\documentstyle[12pt,epsf]{article}
\def\DO{D\O ~}
\def\rbar{\langle r \rangle }
\def\I{{\rm I}}
\def\II{{\rm II}}
\def\III{{\rm III}}

\def\half{\frac{1}{2}}
\def\Jetrad{J{\sc etrad}}
\def\rsep{R_{\rm sep}}
\def\npscale{C}
\begin{document}
\begin{titlepage}
\begin{flushright}
hep-ph/9706210
FERMILAB--Pub--97/163--T\\
Saclay--SPhT--T97/54\\
DTP/97/42\\
\end{flushright}
\vspace{1cm}
\begin{center}
{\Large\bf Jet Investigations Using the Radial Moment}\\
\vspace{1cm}
{\large
W.~T.~Giele}\\
\vspace{0.5cm}
{\it
Fermi National Accelerator Laboratory, P.~O.~Box 500,\\
Batavia, IL 60510, U.S.A.} \\
\vspace{1cm}
{\large
E.~W.~N.~ Glover}\\
\vspace{0.5cm}
{\it
Physics Department, University of Durham,\\ Durham DH1~3LE, England} \\
\vspace{0.5cm}
and \\
\vspace{0.5cm}
{\large
David~A.~Kosower}\\
\vspace{0.5cm}
{\it
Service de Physique Th\'eorique, Centre d'Etudes de Saclay,\\
F--91191 Gif-sur-Yvette cedex, France}\\
\vspace{0.5cm}
{\large \today}
\vspace{0.5cm}
\end{center}
\begin{abstract}
We define the radial moment, $\rbar$, for jets produced in
hadron-hadron collisions.  It can be used as a tool for studying,
as a function of the jet transverse energy and pseudorapidity,
radiation {\it within\/} the jet and the quality of a perturbative
description of the jet shape.
We also discuss how
non-perturbative corrections to the jet transverse energy affect $\rbar$.
\end{abstract}

\end{titlepage}

Prospectors for new physics congregrate at the high-energy frontier.
Their searches require
confronting experimental data with theoretical expectations.
Confidence in their claims of new physics presupposes not only
a proper comparison of data to theory, but also a solid understanding
of systematic uncertainties in both theory and experiment.
An interesting case study is the allegation of new physics based on
the high-energy tail of the single-jet inclusive transverse energy
distribution measured by the CDF and \DO detectors at
Fermilab \cite{CDF,D0}.
Data from both experiments agree well with both the theory (EKS \cite{EKS}
for CDF and \Jetrad\ \cite{Jetrad,GGK,GG} for \DO)
and with each other
for jets of transverse energy less than about 200~GeV, while at
higher energies the CDF data appear to be somewhat larger than expected
(see for example \cite{LT}).
A genuine rise above theoretical expectations at large transverse energies
would be the signal of one of a
whole panoply of new physics possibilities.
Confidence in such a claim, however, requires that we first rule out
more prosaic explanations, such as an uncertainty in the distributions of
partons inside the
colliding nucleons \cite{GMRS,CTEQJET}, one of the
non-perturbative
inputs to the theoretical computation.
In order to help resolve the apparent discrepancies between the experiments,
and also to clarify puzzling aspects of the
cross sections measured at lower energies
\cite{CDF630,D0630},
we feel it is important to examine other observable quantities in the same
high transverse-energy events.
For example, Ellis, Khoze and Stirling \cite{antenna} have proposed
studying the
radiation between jets in order to identify the underlying color structure
of the hard scattering.

In this paper, we advocate studying the radiation {\em within} a jet, and
its use as a diagnostic tool for the quality of the perturbative description
of the jet shape as a function of transverse energy.
Both CDF \cite{CDFshape} and \DO \cite{D0shape}
have measured the transverse-energy profile of jets,
where the integrated density as a function of the radius $r$
from the jet axis,
$$
\Psi(r,E_T,\eta) = \frac{\int_0^r\,E_T(r')dr'
\frac{d\sigma}{dE_T d\eta }}{\int_0^R\,E_T(r')dr'\frac{d\sigma}
{dE_T d\eta }}.
$$
In terms of particle (tower) $i$ within the jet and lying at a distance
$r_{i,{\rm jet}}$
from the jet axis, where,
\begin{equation}
r^2_{i,{\rm jet}} = (\Delta \phi_{i,{\rm jet}})^2
+ (\Delta \eta_{i,{\rm jet}})^2,
\end{equation}
we have,
$$
\Psi(r,E_T,\eta) = \frac
{\left\langle \sum_{i, r_{i,{\rm jet}} < r} E_{Ti}(r_{i,{\rm jet}})
\frac{d\sigma}{dE_T d\eta }\right\rangle_{\rm jets}}
{\left\langle
 \sum_{i} E_{Ti}(r_{i,{\rm jet}})
\frac{d\sigma}{dE_T d\eta }\right\rangle_{\rm jets}}.
$$
The summation runs over all particles (towers) in the jet and
$\sum_i E_{Ti}(r_{i,{\rm jet}}) = E_{T{\rm jet}}$.
These profiles
have also been discussed from the theoretical point of view by Ellis,
Kunzst and Soper \cite{EKSshape1,EKSshape2} and more recently by Klasen and
Kramer \cite{KK}.

While one can
make a qualitative or semi-quantitative comparison with theoretical
pre\-dic\-tions,\footnote{From the perturbative point of view $\Psi$ suffers
from large collinear logarithms for small $r$.}
the language of jet profiles does not lend itself readily
to the extraction of fundamental quantities, or to identifying possible
problems with underlying events or the energy calibration of jets.
It would be better to summarize
the jet profiles in a small set of numbers, whose variation with $E_T$
and $\eta$ can then be mapped more easily.  The simplest such number
is the radial moment of the transverse energy distribution within the jet,
which we shall denote $\rbar$,
\begin{equation}
\rbar =
\frac{\left\langle
 \sum_{i} r_{i,{\rm jet}}E_{Ti}(r_{i,{\rm jet}})
       \frac{d\sigma}{dE_T d\eta }\right\rangle_{\rm jets}}
{\left\langle
 \sum_{i} E_{Ti}(r_{i,{\rm jet}})
\frac{d\sigma}{dE_T d\eta }\right\rangle_{\rm jets}}.
\label{eq:rbar}
\end{equation}
In the central region, $E_T \sim E$, and $r \sim \theta$; for
massless partons within
narrow jets, $r E_T$ is then roughly $\theta E \sim |{\bf k}|\sin\theta$,
where $\theta$ is the angle between the parton and the jet axis.  This
quantity is simply the transverse momentum of the parton with respect
to the jet axis. The radial moment can thus be understood
for narrow jets in the central region as the average transverse
momentum with respect to the jet axis, divided by the jet's $E_T$.
In the context of two-jet inclusive distributions, the triply-differential
distribution $d^3\sigma/dE_T d\eta_1 d\eta_2$
we have discussed previously \cite{Triply} extracts as
much {\it global\/} information about the event as possible, from the
viewpoint of perturbative QCD.  The moment $\rbar$ is independent of this
distribution, and its comparison with theoretical predictions may provide
us with additional information.
As we shall discuss below, the
moment is sensitive to the size of the underlying event as well as to the
experimental jet algorithm.

The moment
vanishes in a lowest-order calculation of jet cross sections, since
jets at that order are approximated by lone partons; a next-to-leading
order calculation of jet cross sections thus gives the leading non-vanishing
calculation of the moment.
Using the \Jetrad\ implementation \cite{Jetrad,GGK,GG} of ${\cal
O}(\alpha_s^3)$
parton scattering processes \cite{ES}, it is straightforward to evaluate
$\rbar$
for an arbitrary jet algorithm.
The dependence of the moment on
$E_T$ using a perturbative implementation of the Snowmass algorithm with
$R=0.7$ is displayed in fig.~1.
To mimic the \DO acceptance,
the jets have been restricted to the central rapidity region,
$|\eta| < 0.2$.
For reference, we use the CTEQ4M parton distribution set together with
a renormalization/factorization scale $\mu = 0.5 E_T$.

\begin{figure}\vspace{8cm}
\includegraphics{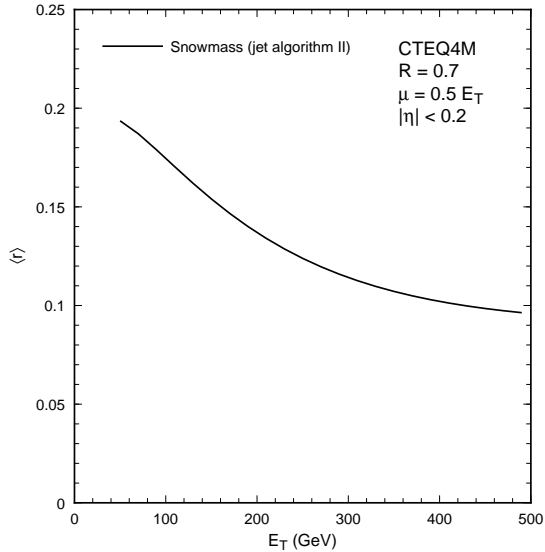}
\caption[]{The moment $\rbar$ as a function of the jet transverse
energy at $\sqrt{s}=1800$~GeV using the CTEQ4M parton distributions,
for jets with $|\eta|<0.2$ reconstructed using a perturbative implementation of
the Snowmass algorithm (jet algorithm II defined in eq.~7) for the jet cone
size $R = 0.7$.
The renormalization and factorization scales have been chosen to be equal
to $0.5 E_T$.
These parameters form our baseline choice.}
\label{fig:example}
\end{figure}

For narrow central jets, we can model the radiation pattern we expect
using the collinear approximation to the matrix elements.
We find,
\begin{equation}
\rbar \sim g^2(\mu)\int r \frac{ds dz}{16\pi^2} \frac{1}{s} P(z)\,,
\end{equation}
where $P(z)$ is an Altarelli-Parisi splitting function.
The dependence on the hard cross section, including the
dependence on the parton distribution
functions, disappears; only the dependence
on $\alpha_s$ remains.  With $\theta_i$ the angle between the jet axis
and parton $i$,
\begin{eqnarray}
s &=& 2 z (1-z) E^2 (1-\cos(\theta_1+\theta_2))\nonumber \\
&\sim& z (1-z) E^2 (\theta_1+\theta_2)^2.
\end{eqnarray}
Balancing of transverse momentum within the jet gives the additional relation,
\begin{displaymath}
z\theta_1 = (1-z) \theta_2.
\end{displaymath}
With the change of variables, $r_1=z\theta_1$, $r_2 = (1-z)\theta_2$, we have,
\begin{equation}
\rbar  \sim   \frac{g_s^2(\mu)}{4\pi^2} \left (
\int  dz \,z P(z)\int  d\theta_1
+ \int dz \,(1-z) P(1-z)\int d\theta_2 \right). \nonumber
\end{equation}

The limits in our integral are determined by the clustering in the
jet algorithm we are using.
At next-to-leading order in perturbation theory,
there are two distinct choices for whether or not the two partons will be
clustered to form a jet of cone-size $R$, depending on whether the
parton-parton  or parton-jet distance is constrained to be less than $R$;
\begin{eqnarray}
\I.~~~&&r_{12} < R,\\
\II.~~~&&r_{i,{\rm jet}} < R.
\end{eqnarray}
In this latter algorithm, two partons with equal transverse energy
will combine to form a single jet even if $r_{12} = 2R$.   In this case, the
jet will have all the transverse energy lying on the edge of the jet.
It is unlikely that an experimental jet, made up of transverse energy smeared
over many calorimeter cells will reconstruct such a jet; it is much more likely
to find two (smaller $E_T$ jets), with an overlapping region.
This is impossible to model accurately at next-to-leading order when at most
two partons can merge.  However, it is usually \cite{EKSshape1,EKSshape2,KK}
approximated by the constraint,
\begin{equation}
\III.~~~r_{i,{\rm jet}} < R\qquad {\rm and}\qquad r_{12} < \rsep R,
\end{equation}
where $1 < \rsep < 2$.
When $\rsep = 1$, we find jet algorithm $\I$, while
for $\rsep=2$ we obtain algorithm $\II$.
Even after choosing the clustering criterion,
there still remains a choice of
recombination scheme.  The most common choice is the
Snowmass recombination method \cite{Snowmass} originally designed to make
semi-analytic theoretical calculations feasible.
Since this is a reasonable approximation to the CDF and \DO\ recombination
schemes in the central region\footnote{
As detailed in ref.~\cite{KG}, the \DO recombination scheme
overestimates the energy of a jet in the forward region, and hence
leads to distortions in the shape, rendering it unsuitable for
the studies envisaged here.} we choose to adopt this scheme for the calculation
of $\rbar$ at ${\cal O}(\alpha_s^3)$.

For the jet cluster  algorithm $\III$, and assuming $\theta_1 > \theta_2$,
we have the constraints $\theta_1 < R$ and $\theta_1 < (1-z) \rsep R$.
This translates into the bounds,
$\theta_1 < R$ and $\theta_2 < zR/(1-z)$ for $z < (\rsep-1)/\rsep = x$
while $\theta_1 < (1-z)\rsep R$ and
for $(\rsep-1)/\rsep < z < 1/2$.
Summing over both partons (and including a factor of 2
for the case $\theta_2 > \theta_1$), yields,
\begin{eqnarray}
\rbar_{\III} &=& \frac{\alpha_s(\mu)}{\pi}
\left( \int_0^{x} dz \,z P(z)\int_0^R d\theta_1
+ \int_x^{\half} dz \,(z) P(z)\int_0^{(1-z)\rsep R} d\theta_1\right .
\nonumber \\
&&
\quad \left .
+ \int_x^{\half}  dz (1-z) P(1-z)\int_0^{z\rsep R} d\theta_2
+ \int_{0}^x dz \,(1-z) P(1-z)\int_0^{zR/(1-z)} d\theta_2\right )
\,\nonumber \\
&\equiv & \frac{\alpha_s(\mu)}{\pi} R {\cal P}^{\III},
\end{eqnarray}
There are two possibilities to consider: where the parent parton is
(a) a quark or (b) a gluon.
Integrating the appropriate
splitting functions, we obtain,
\begin{eqnarray}
{\cal P}^{\III}_{gluon} &=& \frac{
(60\ln(\rsep)+5+60/\rsep+10/\rsep^3-3/\rsep^4-30/\rsep^2-21\rsep)
N}{15} \nonumber \\
&& +
\frac{(20-30/\rsep+30/\rsep^2-20/\rsep^3+6/\rsep^4-3\rsep)
n_f}{30},\nonumber \\
{\cal P}^{\III}_{quark} &=&
\frac{(8\ln(\rsep)+2+6/\rsep-3\rsep-2/\rsep^2)(N^2-1)}{4N}.
\end{eqnarray}

Inserting the numerical values $N=3$ and $n_f = 5$ and fixing
$\rsep = 1$ we obtain the radial moments for jet cluster
criterion $\I$,
\begin{eqnarray}
{\cal P}^{\I}_{gluon} &=& \frac{14N + n_f}{10} = 4.7,\nonumber \\
{\cal P}^{\I}_{quark} &=& \frac{3(N^2-1)}{4N} = 2,
\end{eqnarray}
while for $\rsep = 2$ and jet cluster criterion $\II$, we
find,
\begin{eqnarray}
{\cal P}^{\II}_{gluon} &=& \frac{(192\ln(2)-43)N + 7n_f}{48} = 6.36,\nonumber
\\
{\cal P}^{\II}_{quark} &=& \frac{(16\ln(2)-3)(N^2-1)}{8N}=2.70.
\end{eqnarray}
As expected, this gives rise to somewhat fatter jets.
However,
in both cases, ${\cal P}_{gluon}/{\cal P}_{quark}\sim 2.35$ so that the
relative fatness of `gluon' and `quark' jets is preserved.
Furthermore, \break
${\cal P}^{\II}_{gluon}/{\cal P}^{\I}_{gluon}\sim
{\cal P}^{\II}_{quark}/{\cal P}^{\I}_{quark} \sim  1.35.$
Other choices for $\rsep$ yield values smoothly dispersed between these
two
extremes but preserving the relative fatness of the types of jets.

\begin{figure}\vspace{8cm}
\includegraphics{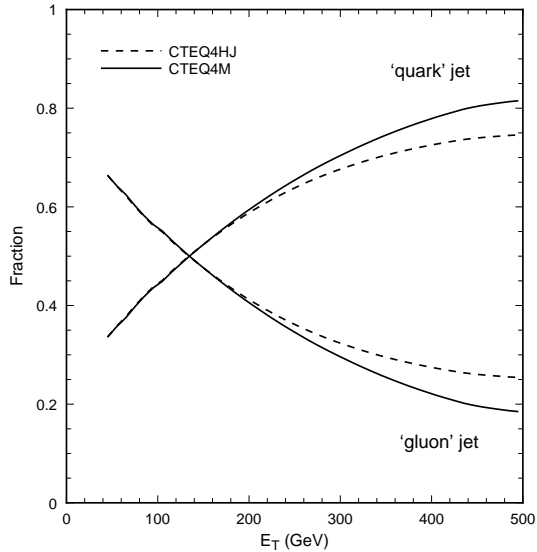}
\caption[]{The fraction of `quark' and `gluon' jets
for $|\eta| < 0.2$ and $\sqrt{s} = 1800$~GeV as
a function of the jet transverse energy
for the CTEQ4M and CTEQ4HJ parton density functions.}
\label{fig:qgfrac}
\end{figure}

\begin{figure}\vspace{8cm}
\includegraphics{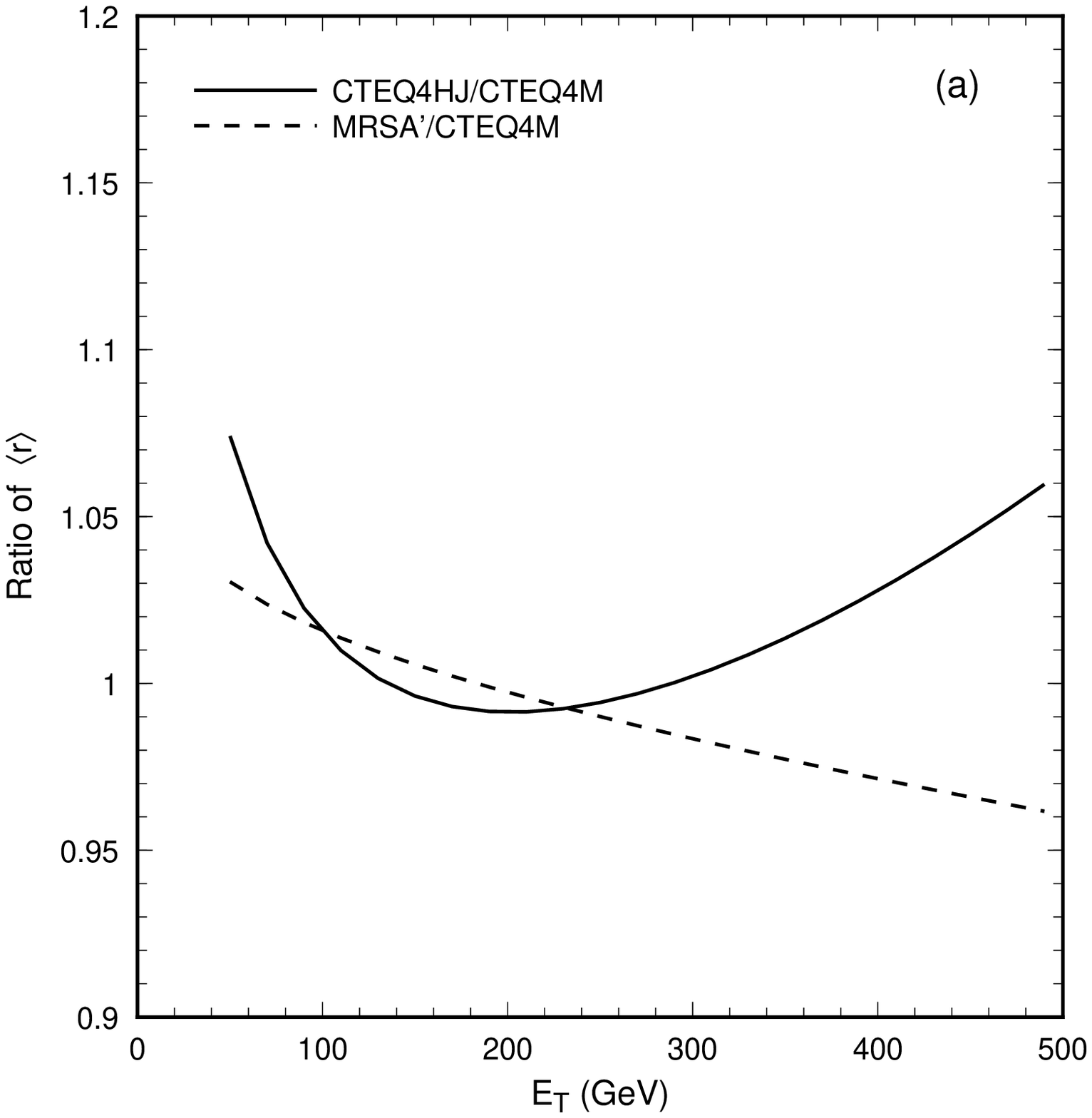}
\includegraphics{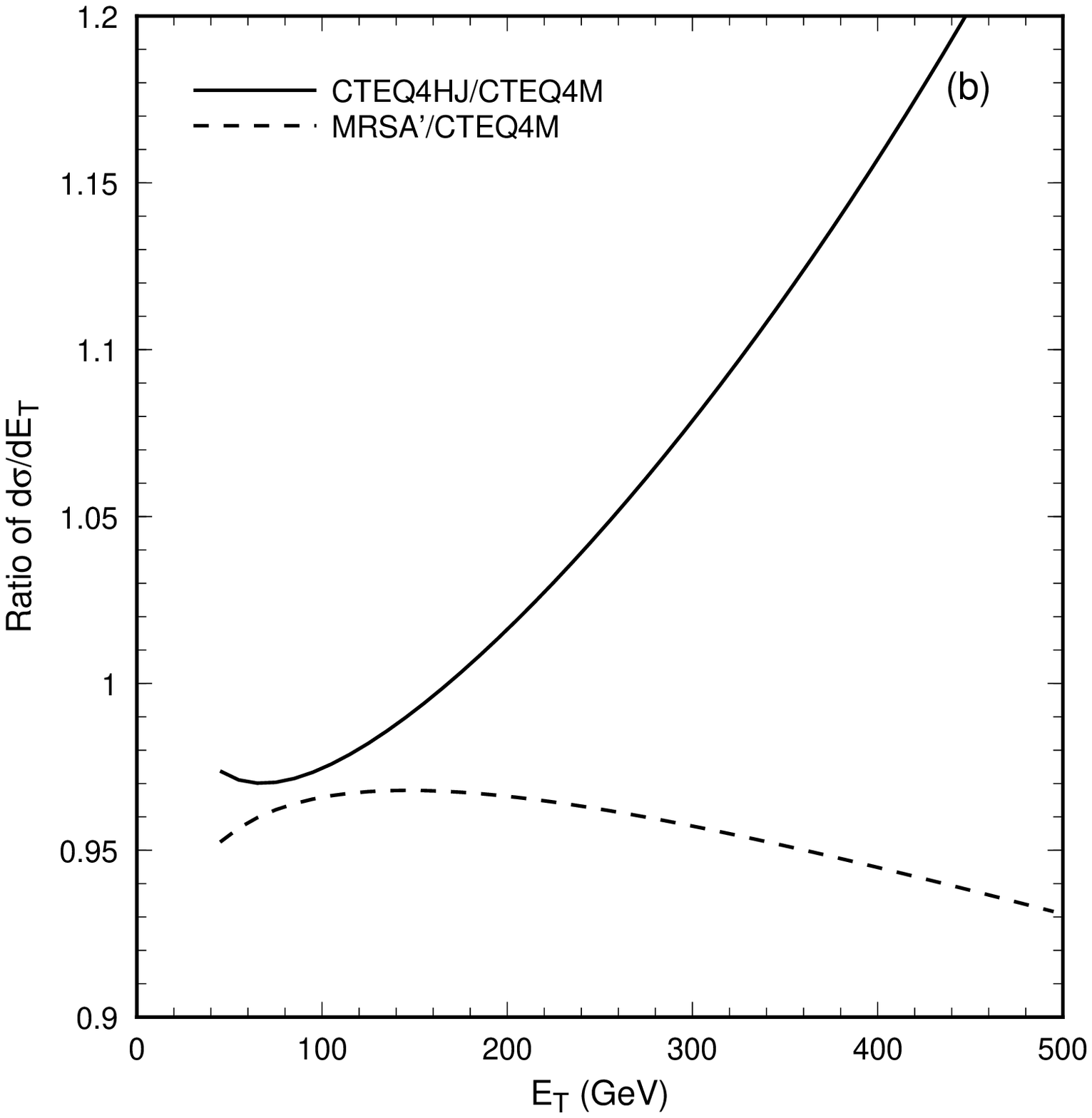}
\caption[]{(a) The ratio of $\rbar$ for the CTEQ4HJ and MRSA${}'$
$(\alpha_s(M_Z) = 0.115)$
distribution sets to that for the CTEQ4M set, as a function of the jet
transverse energy.
(b) The ratio of the single-jet inclusive cross section
computed for the CTEQ4HJ and MRSA${}'$ $(\alpha_s(M_Z) = 0.115)$
distribution sets to that for the CTEQ4M set.
In both cases the renormalization and factorization scales are equal
to $0.5 E_T$ and $|\eta| < 0.2$ and $R=0.7$.}
\label{fig:pdfIndependence}
\end{figure}

As the jet-defining radius $R$ increases, the collinear approximation
gets worse, but none\-the\-less we expect the general features of this
simple model to survive: the moment should be roughly a constant
times $\alpha_s(\mu)$, where $\mu$ is a scale characterizing the
jet.  In particular, it should be nearly independent of the parton
distribution function.
However, the parton distributions do play an important role.
As we increase the hardness of the scattering we sample
different types of jet.
The matrix elements for the different subprocesses are (within color factors)
rather similar for scattering at $\sim 90^{\circ}$,
and the dominant effect is due to the variation of the
quark and gluon parton densities with $x$.
This is illustrated in fig.~\ref{fig:qgfrac} where we show the
fraction of `quark' and `gluon' jets at
$\sqrt{s} = 1800$~GeV as a function of the jet transverse energy
for the CTEQ4M parton distributions \cite{CTEQ4}.
While `gluon' jets dominate at $E_T < 120$~GeV,
as the quark density functions become more important, the fraction of `gluon'
jets diminishes.
Different parton distributions give quite different results.
In particular, the CTEQ fit to the supposed
excess of jets high transverse momentum
CTEQ4HJ \cite{CTEQ4} containing an enhanced gluon at large $x$
generates a larger fraction of
`gluon' jets at high $E_T$.\footnote{
Note that in order to fit the jet excess at high $E_T$, the inclusive jet
cross section also increased as can be seen in fig.~\ref{fig:pdfIndependence}.}

This variation of $\rbar$ with the choice of distribution function set
is  contrasted in  fig.~\ref{fig:pdfIndependence}  with the relatively
greater  sensitivity of  the inclusive-jet spectrum.    To make a fair
comparison, we use the CTEQ4 fit  with an enhanced  gluon at large $x$
that  describes the  single jet  inclusive  distribution well  at high
$E_T$, CTEQ4HJ \cite{CTEQ4},   and   a  `normal'   fit  by the     MRS
collaboration  \cite{MRSAP} with  a similar  value of $\alpha_s(M_Z)$.
The relative insensitivity to parton distribution functions makes this
moment a useful tool for measuring the strong coupling constant.

\begin{figure}\vspace{8cm}
\includegraphics{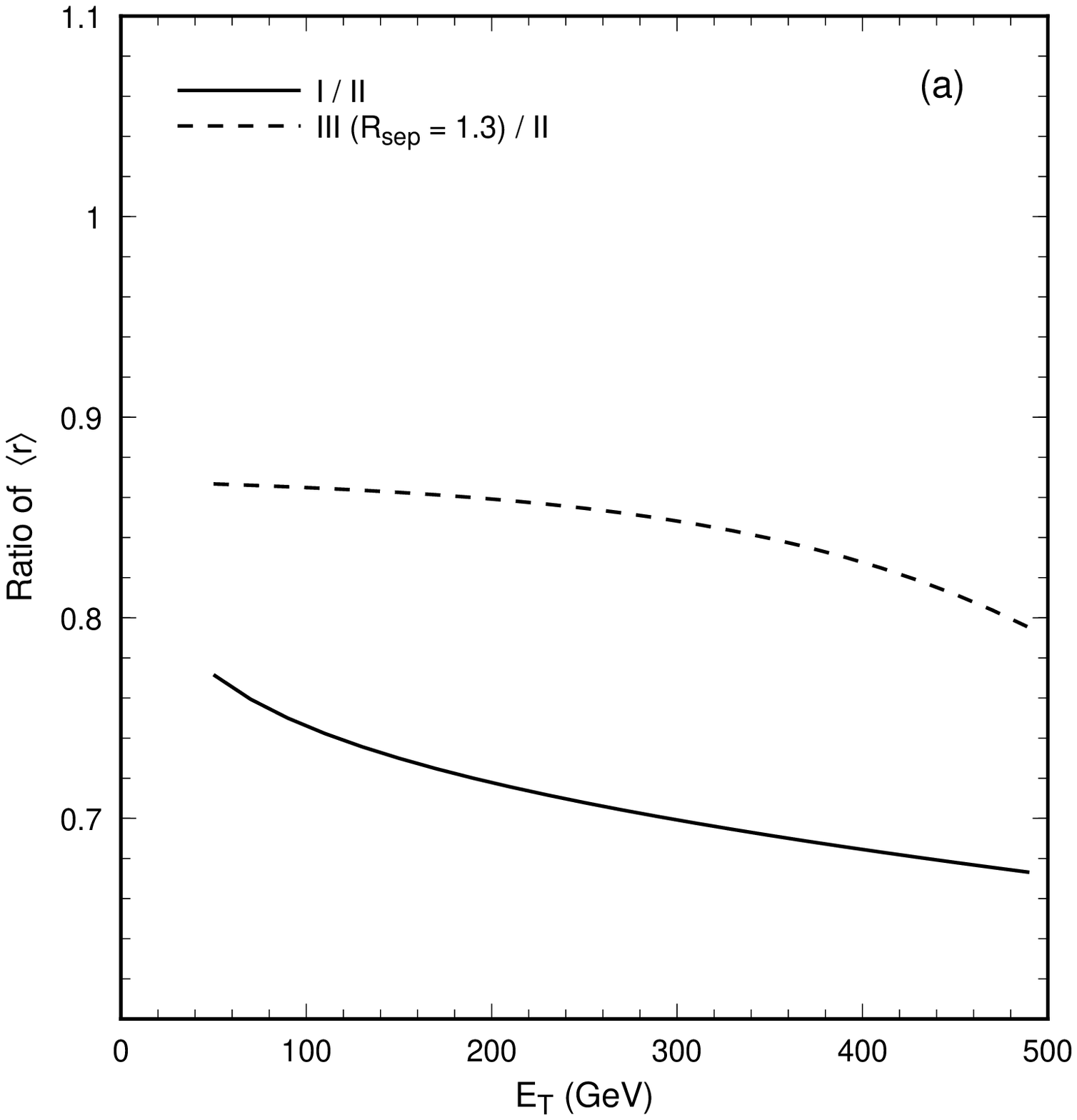}
\includegraphics{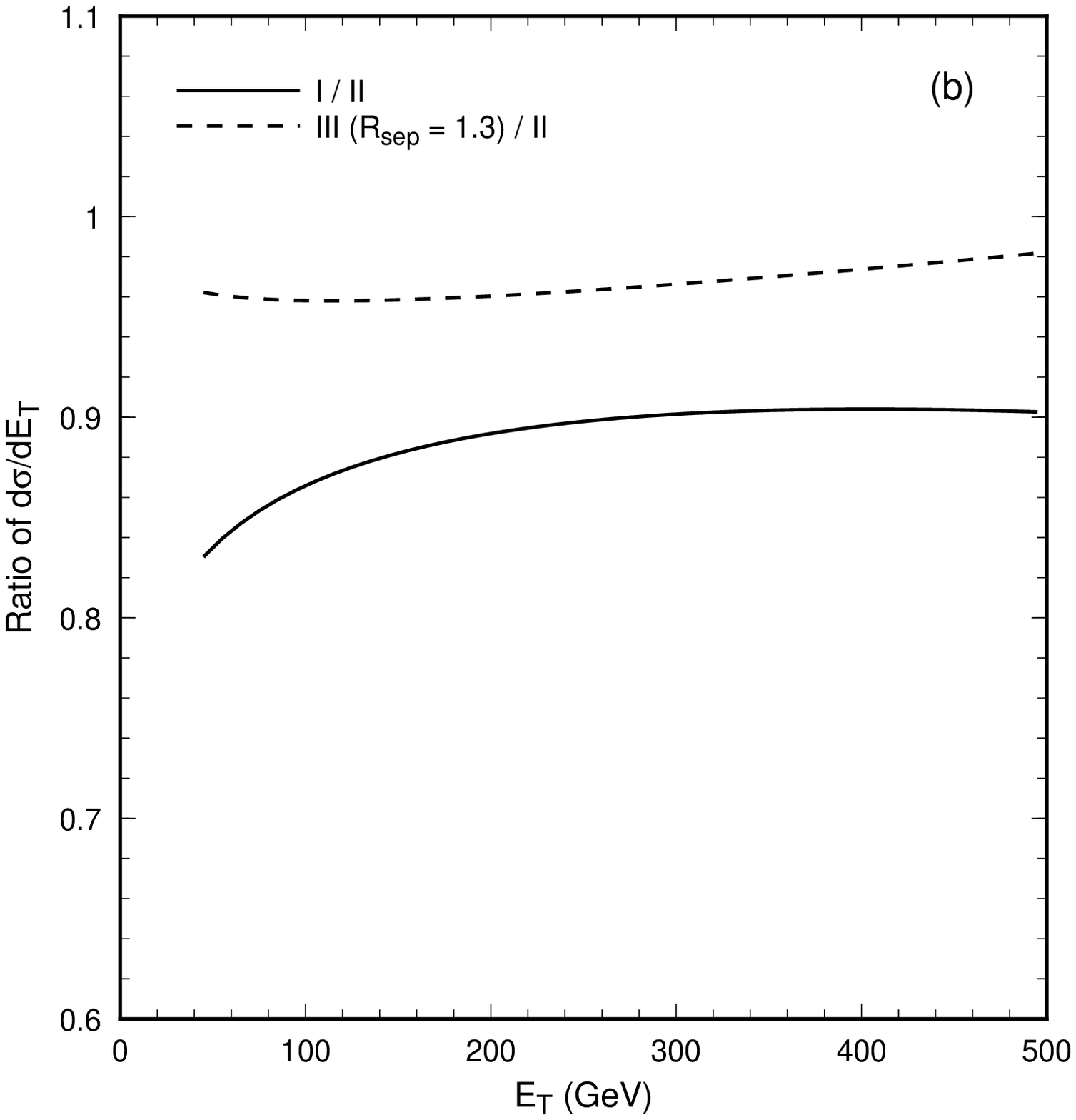}
\caption[]{(a) The ratio of $\rbar$ for the jet algorithms defined in the text
as function of the jet transverse energy.
(b) The ratio of the single-jet inclusive cross section for the same jet
algorithms.
In both cases, CTEQ4M parton distributions have been used
with renormalization and factorization scales equal
to $0.5 E_T$ and $|\eta| < 0.2$ and $R=0.7$.}
\label{fig:algorithm}
\end{figure}

The radial moment is also
somewhat more sensitive to the jet algorithm than
the single-jet inclusive cross section, as illustrated in
fig.~\ref{fig:algorithm} for jet algorithms $\I$
and $\III$ with $R = 0.7$ and $\rsep=1.3$ relative to algorithm $\II$.
Once again, the jets are constrained to lie centrally, $|\eta| < 0.2$.
We see that both the normalization, and to a lesser extent,
the rate of decrease of $\rbar$ vary more than the single-jet
$E_T$ spectrum under changes of the jet algorithm.

\begin{figure}\vspace{8cm}
\includegraphics{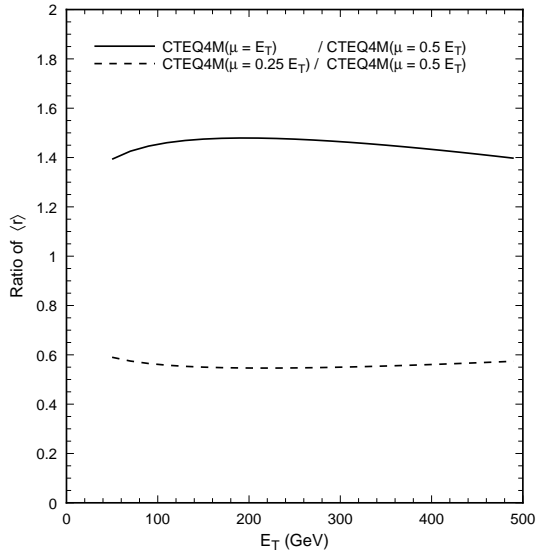}
\caption[]{The radial moment $\rbar$ for renormalization/factorization scale
 $\mu = \lambda E_T$ relative to that for $\mu = E_T$
for central jets, $|\eta | < 0.2$ using the
cluster criterion $\II$ and the Snowmass recombination procedure.
The solid (dashed) lines shown the change for $\lambda = 0.5$ (2.0).
In all cases, the MRSA${}'$ parton distributions have been utilized.}
\label{fig:scale}
\end{figure}

Of course, the calculations presented
here will suffer a significant renormalization-scale dependence, as they
are leading-order calculations of the radial moment; a measurement of
$\alpha_s$ would require a next-to-leading order calculation of this
quantity.
To estimate the scale uncertainty, we compute
the ratio of $\rbar$ computed with
$\mu = 0.25 E_T$  and $\mu = E_T$ relative
to the same quantity at our reference scale choice
$\mu = 0.5 E_T$ for jet clustering algorithm $\II$.
We see  in fig.~\ref{fig:scale}, the scale variation is still sizeable
and affects the normalization considerably.
On the other hand, varying the scale hardly changes the shape.

Armed with the perturbative predictions, we can ask whether the
radial moment is sensitive to non-perturbative contributions, either to the
jet energy or to the pattern of radiation within the jet.
These may arise from various sources, such as power ($1/Q$) corrections
to the hard scattering process, spectator interactions, or
soft hadrons in the proton/antiproton remnants spilling into the jet cone.
Another source of uncertainty is detector-related effects such as
the jet energy-scale uncertainty.
For example, the debris of the hard scattering tends to form a roughly uniform
distribution in pseudorapidity and azimuth. Shower models are used to estimate
this underlying event and thereby correct the jet energy.
These have been extensively tuned at moderate energies where the data are
plentiful.
However, if the underlying event was not properly modelled as a function of
$E_T$,
a residual jet energy correction could occur.

In the simplest string-like hadronization model~\cite{Tube}, we expect
to find a contribution of order $\npscale\sim 1$~GeV to the transverse
momentum $p_t$ with  respect to the jet axis  in a central jet.  Since
the   average  $p_t$   is roughly   $\rbar    E_T$,  this  implies   a
non-perturbative  correction of order $C/(\rbar  E_T)$.  We expect the
corrections to the   radial moment to  be   of the  same order.    The
underlying  event is also expected to  contribute  a $1/Q$ correction.
If  we  assume that  the energy is  deposited  uniformly,  we expect a
contribution to $\rbar$ of order
\begin{equation}
\frac{\pi E_{u}}{E_T}\left(\frac{2R}{3}-\rbar\right)\,,
\end{equation}
where   $E_u$ is the underlying   event transverse energy density (per
unit   rapidity per radian).  In  minimum-bias   events, this is $\sim
0.55\pm 0.1$~GeV according to ref.~\cite{AbbottsThesis}.  Some of this
is   subtracted   as  part of  the     experimental analysis.
Since $\rbar$  varies from $\sim 0.25$ to $\sim
0.07$, this  correction should  anyway be much  smaller  than that    due to
hadronization.

A determination of $\alpha_s$, which is less sensitive to the parton
distributions than
other determinations at hadron colliders, and the possible identification
of detector and/or physics based non-perturbative contributions are
both interesting to pursue. This gives the jet-shape analysis a well
defined physics goal, and motivates the determination of jet shapes
over a large range of transverse energies. While a sensible analysis
requires at least the next-to-leading order contributions
to the jet shape, we can already confront the published CDF and \DO data on
the quantity $\Psi(r,E_T,\eta)$ defined in eq.~1 with the leading-order
calculations. This allows us to explore the potential utility of this
measurement,  and indicate expected uncertainties and
problems.

The current published data on the jet shape from both CDF and \DO
are very limited. The CDF collaboration published results based on
4.2~pb$^{-1}$ with only three bins of transverse energy, while
\DO used 13~pb$^{-1}$ with four bins in transverse energy.
A full analysis of
current data would include on the order of 100 pb$^{-1}$
from each collaboration, preferably
using the same binning in $E_T$ as used
in the one-jet inclusive transverse-energy distribution.
Substantial experimental improvements to the results presented here
are thus possible. In addition, we expect the next-to-leading order
theoretical corrections to be available soon. The latter will enable
us to extract the next-to-leading order $\alpha_s$ and determine
the uncertainty on the radial moment, and will be crucial
in understanding the magnitude of non-perturbative effects.

We must first determine the radial moment $\rbar$
from the published results for the quantity $\Psi$. In an analysis
along the lines suggested in this paper one would of course determine the
radial moment as a function of transverse energy directly from the data
without resorting to extracting the explicit shapes. This would
reduce the experimental uncertainties significantly, especially
the systematic uncertainties. For now, we have to reconstruct
the radial moment from the quantity $\Psi$:
\begin{equation}
\langle r(E_T,\eta)\rangle = \int_0^R r\frac{\partial \Psi(r,E_T,\eta)}
{\partial r} d\,r\ .
\end{equation}
The values of $\rbar$ extracted from the published data are shown in
Table~1.
Both experiments use $R=1$, but the allowed range of rapidity for the jets is
different.
CDF consider jets in the range $0.1 < |\eta| < 0.7$ while \DO have a tighter
restiction, $|\eta| < 0.2$.
\begin{table}\begin{center}
\begin{tabular}{|c|c|c|c|c|}\hline
$E_T$-bin (GeV) & $\langle E_T\rangle$ (GeV) & experiment  &
$\langle r\rangle|_{\mbox{experiment}}$
\\ \hline
40--60    & 45  & CDF & 0.31 $\pm$ 0.02 \\
45--70    & 53  & \DO  & 0.29 $\pm$ 0.01 \\
65--90    & 75  & CDF & 0.24 $\pm$ 0.02 \\
70--105   & 81  & \DO  & 0.23 $\pm$ 0.01 \\
95--120   & 100 & CDF & 0.22 $\pm$ 0.02 \\
105--140  & 118 & \DO  & 0.19 $\pm$ 0.01 \\
140--900  & 166 & \DO  & 0.17 $\pm$ 0.01 \\
\hline\hline \end{tabular}
\caption[]{The \DO and CDF experimental results for $\rbar$ extracted
from the published jet shapes.}
\end{center}\end{table}

For the theoretical prediction, we could repeat the experimental extraction,
by first determining
$\Psi$ using the same binning as the experiments and extract
$\rbar$ from this quantity.
This will simulate any effects on $\rbar$ due to
binning and functional parametrization.
However, these effects are in practice small and
instead
we determine $\rbar$ directly as was done in fig.~\ref{fig:example},
For the theoretical prediction,
we use the CTEQ4M parton distribution set, with $\alpha_s(M_Z) = 0.116$.
The renormalization and factorization scales were chosen to be half the
jet transverse energy.
To match the experimental cuts, we use $R=1$ and
$\rsep = 1.3$ was used to simulate the jet splitting and merging effects,
which are
not modelled at this order in perturbation theory.

\begin{figure}\vspace{8cm}
\includegraphics{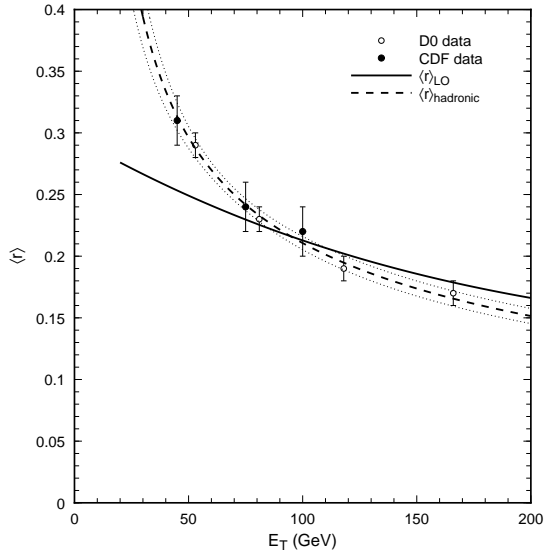}
\caption[]{The extracted values for $\rbar$ for \DO (open circles) and
CDF (solid circles).   The theoretical prediction for $|\eta| < 0.2$, CT 4M
parton distributions, $R=1$ and $\rsep = 1.3$ and $\mu= 0.5 E_T$ is shown as a
solid line. The fitted theoretical result with an additional non-perturbative
component as described in eq.~\ref{eq:np} and $\npscale=1.09$ and $K=0.56$ is
given by the dashed line;
the dotted lines indicate the 1$\sigma$ variations in the fit.}
\label{fig:data}
\end{figure}

The extracted experimental values for $\rbar$ are shown in
fig.~\ref{fig:data} along with the theoretical prediction
as a function of $E_T$.
We see that the CDF and \DO points are compatible
with each other (despite the fact that the experiments use slightly
different rapidity cuts on the jet).  On the other hand, the leading-order
predictions cannot describe the data, as its $E_T$ dependence is substantially
different from that of the data.  As was already demonstrated in
fig.~\ref{fig:scale}
while the normalization of the leading-order prediction has a strong dependence
on the choice of the renormalization/factorization scale, the
shape is mostly independent of this choice.  Thus, although
we assume that higher-order corrections will modify the normalization, it
is reasonable to assume that they will not modify the shape substantially.

This suggests that the explanation of the shape of the data lies in
hadronization effects (i.e. power corrections).
To test this hypothesis, we will assume
that the perturbative prediction $\rbar_{\it perturbative}$
is a constant $K$ times the leading-order prediction $\rbar_{LO}$. We
model the hadronization effects as described earlier, so that
\begin{eqnarray}
\rbar_{\it hadronic} &=&
\rbar_{\it partonic} + \frac{\npscale}{\rbar_{\it partonic}E_T^{\rm jet}}
\nonumber \\
&=& K\rbar_{\rm LO} + \frac{\npscale}{K\rbar_{\rm LO}E_T^{\rm jet}}
\label{eq:np}
\end{eqnarray}
where the non-perturbative scale $\npscale$ should be close to 1 GeV.
We then perform a fit of this model to the data varying both the
$K$-factor and the hadronization scale $\npscale$.
The fit result with a hadronization scale $\npscale=1.09\pm 0.05$~GeV
and $K = 0.56 \pm 0.11$ is shown in
fig.~\ref{fig:data}.
Other choices of renormalization scale give roughly the same value of
$\npscale$.
This is not surprising because
$\npscale$ changes the shape of the curve while
the renormalization scale affects primarily the normalization.
This shows that the above model
might explain the observed differences between the leading-order predictions
and the data. The $K$-factor has a strong dependence
on the renormalization scale, as expected.

Only a next-to-leading order calculation of
$\rbar$ can confirm that power corrections, rather than higher-order
corrections, are responsible for the observed behavior of the data.
Such a calculation requires the one-loop five-parton matrix
elements~\cite{FiveParton}
as well as the six-parton tree-level ones~\cite{SixParton}, all
of which are available in the literature.
With such NLO predictions, one could extract both an NLO
$\alpha_s$ (sensitive to the
normalization and the high-$E_T$ dependence of $\rbar$)
and the hadronization scale $\npscale$ (sensitive to the low-$E_T$
dependence of $\rbar$).
In order to perform such an extraction, however,
we need a much more extensive measurement
of the $E_T$ dependence of $\rbar$.  With current CDF
and \DO data sets, it should be possible to cover a much larger
$E_T^{\rm jet}$ range (both lower and higher than the current published
results) with a finer binning.  This would enable a study of
power corrections and their impact on the uncertainty in an extracted
$\alpha_s$.

A few remarks are in order.  The current iterative cone algorithm will
be rather problematic in such an NLO calculation (these are the
problems one would encounter in a next-to-next-to-leading order calculation
of jet differential cross sections).
For the studies performed in this paper, we used a phenomenological
parameter $\rsep=1.3$ to model jet splitting and merging.
At the next order in perturbation theory, one should either abandon
this parameter (relying on the theory to model jet splitting and merging),
or else one must introduce an additional $\rsep^{\rm NLO}$.
For the purposes considered here, $\rsep$ is anyway an
especially suspect parameter because it is purely phenomenological and
may well be $E_T$ dependent \cite{KK}.
Indeed, as
we have seen in this paper, jets get narrower as their
energy increases.  With this observation one could
argue that $\rsep$ should in fact decrease with increasing
energy of the jet.
A simple way out of this morass is
to use the $K_T$ jet algorithm \cite{kt,EKSshape2}. It has good
perturbative behaviour and no additional phenomenological
parameters are needed.

The analysis discussed in the present paper suggests that
it should be possible to use the
hadronic dijet system to determine both $\alpha_s$ and the
proton's gluon distribution with great accuracy.  The gluon
distribution would be determined by the triply-differential
distribution, and, once the next-to-leading order calculations are available,
$\alpha_s$ would be determined by
the radial moment.  Only the quark distributions would be taken from
deeply-inelastic scattering data.
Use of the full current data
set would allow a much finer jet $E_T$ binning as well
as much higher jet $E_T$ to be used, reducing
the experimental uncertainties on $\rbar$  significantly
reduced, and thereby reducing the
significance of power corrections.
If the higher-order perturbative corrections are also small,
as suggested by the phenomenological study performed in this paper,
the theoretical uncertainties should be competitively small as well.
We may also expect the high-$E_T$ jet profiles to be especially
sensitive to jet production from new physics, and thus the $E_T$
dependence of the
radial moment should be a good probe of the presence (or absence) of physics
beyond the standard model.


\begin{thebibliography}{99}
\bibitem{CDF}
F. Abe et al, CDF Collaboration, Phys. Rev. Lett. {\bf 77}, 438 (1996).
\bibitem{D0} G. C. Blazey for the \DO Collaboration,
Proceedings 31st Rencontres de Moriond: "QCD and High-energy Hadronic
Interactions",
Les Arcs, France, (1996), p 155.
\bibitem{EKS} S.D. Ellis, Z. Kunszt and D.E. Soper,
Phys. Rev. Lett. {\bf 62}, 726 (1989);
Phys. Rev. {\bf D40}, 2188 (1989);
Phys. Rev. Lett. {\bf 64}, 2121 (1990).
\bibitem{Jetrad} W. T. Giele, E. W. N. Glover, and D. A. Kosower,
 Phys.\ Rev.\ Lett.\ 73:2019 (1994) [hep-ph/9403347]
\bibitem{GGK} W. T. Giele, E. W. N. Glover and D. A. Kosower,
        Nucl. Phys. {\bf B403}, 633 (1993).
\bibitem{GG} W. T. Giele and E. W. N. Glover,
        Phys. Rev. {\bf D46}, 1980  (1992).
\bibitem{LT} H. L. Lai and W.-K. Tung,
`Comparison of CDF and \DO Inclusive Jet Cross Sections',
 hep-ph/9605269.
\bibitem{GMRS} E. W. N. Glover,
A. D. Martin, R. G. Roberts and W. J. Stirling,
        Phys. Lett. {\bf B381}, 353 (1996).
\bibitem{CTEQJET}
J. Huston et al, CTEQ Collaboration, Phys. Rev. Lett. {\bf 77}, 444 (1996).
[hep-ph/9511386]
\bibitem{antenna} J. Ellis, V.A. Khoze and W.J. Stirling,
`Hadronic Antenna Patterns to Distinguish Production Mechanisms for Large-$E_T$
Jets',
hep-ph/9608486.
\bibitem{CDF630} T.~Devlin for the CDF Collaboration,
XXVIIIth International Conference on High Energy Physics, Warsaw, Poland,
July 1996.
\bibitem{D0630} J.~Krane for the D0 Collaboration,
DPF96; Annual Divisional Meeting of the Division of Particles and
Fields of the American Physical Society, Minnesota, U.S.A.,  August 1996.
\bibitem{CDFshape}
F. Abe et al, CDF Collaboration, Phys. Rev. Lett. {\bf 70}, 713 (1993).
\bibitem{D0shape}
S. Abachi et al, \DO Collaboration, Phys. Lett. {\bf B357}, 500 (1995).
\bibitem{EKSshape1} S.D. Ellis, Z. Kunszt and D.E. Soper,
Phys. Rev. Lett. {\bf 69}, 3615 (1992).
\bibitem{EKSshape2} S.D. Ellis and D.E. Soper,
Phys. Rev. {\bf D48}, 3160 (1993).
\bibitem{KK} M. Klasen and G. Kramer,
`Jet Shapes in $ep$ and $p\overline p$ Collisons in NLO QCD',
hep-ph/9701247.
\bibitem{Triply} W. T. Giele, E. W. N. Glover, and D. A. Kosower,
         Phys.\ Rev.\ D52:1486 (1995) [hep-ph/9412338]
\bibitem{ES} R. K. Ellis and J. Sexton,
        Nucl. Phys. {\bf B269}, 445 (1986).
\bibitem{Snowmass} S. D. Ellis, J. Huth, N. Wainer,
K. Meier, N. Hadley, D. Soper, and M. Greco, in {\it Research Directions
for the Decade\/}, Proceedings of the Summer Study, Snowmass,
Colorado, 1990, ed.\ E. L. Berger (World Scientific, Singapore, 1992), p134.
\bibitem{KG}
E.W.N. Glover and D.A. Kosower, Phys. Lett. {\bf 367B}, 369 (1996).
\bibitem{CTEQ4}
H.L. Lai et al, CTEQ Collaboration, Phys. Rev. {\bf D55}, 1280 (1997).
[hep-ph/9606399]
\bibitem{MRSAP} A. D. Martin, R. G. Roberts and W. J. Stirling,
        Phys. Rev. {\bf D50}, 6734 (1994);
        Phys. Lett. {\bf B354}, 155 (1995).
\bibitem{Tube} R. P. Feynman, {\it Photon-Hadron Interactions\/},
W. A. Benjamin (New York, 1972);\\
B. R. Webber, Lectures at the Summer School on {\it Hadronic Aspects
of Collider Physics\/}, Zuoz, Switzerland, August 1994 [hep-ph/9411384].
\bibitem{AbbottsThesis} B.K.  Abbott, Ph.D. thesis, Purdue University,
UMI-95-23307 (1994).
\bibitem{FiveParton}
Z. Bern, L. Dixon, and D. A. Kosower,
    Phys.\ Rev.\ Lett.\ 70:2677 (1993) [hep-ph/9302280];\\
Z. Bern, L. Dixon, and D. A. Kosower, Nucl.\ Phys.\ B437:259 (1995)
[hep-ph/9409393];\\
Z. Kunszt, A. Signer, Z. Trocsanyi, Phys.\ Lett.\ B336:529 (1994)
[hep-ph/9405386].
\bibitem{SixParton} see M. L. Mangano and S.J. Parke, Phys. Reports, {\bf 200},
301 (1991) and references therein.
\bibitem{kt} Yu.L.~Dokshitzer, Contribution to the Workshop on Jets
at LEP and HERA, J. Phys. {\bf G17} (1991) 1441;\
S. Catani, Yu.L.~Dokshitzer, M.H. Seymour  and B.R.Webber,
Nucl. Phys. {\bf B406}, 187 (1993).
\end{thebibliography}
\end{document}